%% file: cells.tex
\pdfoutput=1
\documentclass[sigconf]{acmart}
\acmConference[ESEC/FSE 2018]{Submitted to 11th Joint Meeting of the European Software Engineering Conference and the ACM SIGSOFT Symposium on the Foundations of Software Engineering}{4--9 November, 2017}{Florida, United States}
\renewcommand\footnotetextcopyrightpermission[1]{} 
\usepackage{enumitem}
\setlist{nolistsep}
\usepackage{bigstrut}
\usepackage{amsmath}
\usepackage{booktabs} 
\usepackage{algorithmicx} 
\usepackage[linesnumbered,ruled,vlined]{algorithm2e}
\SetKwInOut{Parameter}{Parameter}
\SetKwInOut{Require}{Require Func}
\SetKwInOut{Input}{Input}
\SetKwInOut{Output}{Output}
\usepackage{algpseudocode}
\usepackage{multirow}
\usepackage{tabu}
\usepackage{enumitem} 
\usepackage{tabularx}
\usepackage{censor} 
\usepackage{subcaption}
\usepackage[framemethod=tikz]{mdframed}
\usetikzlibrary{shadows}
\usepackage{graphics}
\newmdenv[
tikzsetting= {fill=gray!10},
linewidth=1pt,
roundcorner=2pt, 
shadow=false
]{myshadowbox}

\newcolumntype{s}{>{\centering \arraybackslash \hsize=.5\hsize}X}  

\definecolor{pink}{rgb}{0.94,0.2,0.3}
\newcommand{\quart}[4]{\begin{picture}(80,4)
	{\color{pink}\put(#3,2){\circle*{4}}\put(#1,2){\line(1,0){#2}}}\end{picture}}

\usepackage{listings}

\definecolor{MyDarkBlue}{rgb}{0,0.08,0.45} 
\newcommand{\tion}[1]{\S\ref{sect:#1}}
\newcommand{\fig}[1]{Figure~\ref{fig:#1}}
\newcommand{\tab}[1]{Table~\ref{tab:#1}}

\newcommand{\quartex}[3]{
\begin{picture}(13,6)
    {
        \color{black}
        \put(#3,3)
        {\circle*{4}}
        \put(#1,3)
        {\line(1,0){#2}}
    }
\end{picture}
}

\usepackage[framemethod=tikz]{mdframed}
\usepackage{lipsum}
\usetikzlibrary{shadows}
\usepackage{mathtools}
\usepackage{balance}
\usepackage{graphics}

\usepackage{color,colortbl}
\usepackage{xcolor}
\definecolor{lightgray}{gray}{0.8}

\makeatletter
\let\th@plain\relax
\makeatother

\usepackage[framed]{ntheorem}
\usepackage{framed}
\usepackage{tikz}
\usetikzlibrary{shadows}

\definecolor{Gray}{rgb}{0.88,1,1}
\definecolor{lightergray}{gray}{0.95}
\theoremclass{Lesson}
\theoremstyle{break}
\tikzstyle{thmbox} = [rectangle, rounded corners, draw=black,
 fill=Gray,  drop shadow={fill=black, opacity=1}]

\newcommand{\bi}{\begin{itemize}[leftmargin=0.4cm]}
\newcommand{\ei}{\end{itemize}}
\newcommand{\be}{\begin{enumerate}}
\newcommand{\ee}{\end{enumerate}}

\newshadedtheorem{lesson}{Result}

\setcopyright{rightsretained}

\begin{document} 
\setlength{\abovedisplayskip}{3pt}
\setlength{\belowdisplayskip}{3pt}
\title{Transfer Learning with Bellwethers to find Good Configurations}

\author{Vivek Nair, Rahul Krishna, Tim Menzies, and Pooyan Jamshidi*}
\affiliation{%
  \institution{Computer Science, NC State, USA, *Carnegie Mellon University, USA}
}
\email{{vivekaxl, i.m.ralk, tim.menzies, pooyan.jamshidi}@gmail.com}

\begin{abstract}
As software systems grow in complexity, the space of possible configurations grows exponentially. Within this increasing
complexity, developers, maintainers, and users cannot keep track
of the interactions between all the various configuration options.

Finding the optimally performing configuration of a software system for a given setting is challenging. Recent approaches address this challenge by learning performance models based on a sample set of configurations. However, collecting enough data on enough sample configurations
 can be very expensive since each such sample requires configuring, compiling and executing the entire system against a complex test suite.

The central insight of this paper is that choosing a suitable source (a.k.a. ``bellwether'') to learn from, plus
a simple transfer learning scheme will often outperform much more complex transfer learning methods. 
Using this insight, this paper proposes   BEETLE,  a novel bellwether based transfer learning scheme, which can identify a suitable source and use it to find near-optimal configurations of a software system.
BEETLE  significantly reduces the cost (in terms of the number of measurements of sample configuration) to build performance models. We evaluate our approach with 61 scenarios based on 5 software systems and demonstrate that BEETLE  is beneficial in all cases.

This approach offers a new highwater mark in configuring software systems. Specifically,
BEETLE can find configurations that are as good or better as those found by anything else while requiring only $\frac{1}{7}$th of the evaluations needed by the state-of-the-art.
\end{abstract}


\maketitle

\section{Introduction}



Finding the right set of configurations that can achieve the best performance becomes increasingly challenging as the number of configurations increases~\cite{xu2015hey}. 
As
software systems grow in both size and complexity, optimizing a software system to meet the needs of a workload has surpassed the abilities of humans~\cite{bernstein1998asilomar}.  

Much current research has explored this problem, usually by creating accurate performance models that predict performance characteristics.
While this approach is cheaper and more effective  than manual configuration,
it still incurs the expensive of extensive data collection about the software~\cite{guo2013variability, sarkar2015cost, siegmund2012predicting, nair2017faster, nair2017using, nair2018finding, oh2017finding, guo2017data,JC:MASCOTS16}.  This is undesirable,
since this data collection has to be repeated if ever the software is updated
on the    workload of the system changes abruptly. 
In such a scenario, all prior research suffers from the same drawback:
 \begin{quote}
 \textit{These approaches do not learn from previous optimization experiments and must  be rerun whenever the environment of the experiments change.}
\end{quote}
Note our use of the term {\em environment}. This  refers to the external factors influencing the performance of the system such as workload, hardware, version of the software. 


Recent research in performance prediction for configurable systems has shown that {\em transfer learning} can be effective for resolving `cold start' problems---problems for which collecting data is expensive.
Transfer learning typically entails the transfer of information from a selected ``\textit{source}'' software system operating in an environment to learn a predictive model for predicting the performance of the system in the ``\textit{target}'' environment (software system in a different environment), hence enabling a user to reuse measurements. 
This approach has received much attention in the software analytics literature\cite{Nam2013, Nam2015, Jing2015, kocaguneli2011find,kocaguneli2012,turhan09,peters15, krishna18}.

Transfer learning can only be useful in cases where the source environment is similar to the target environment. If the source and the target are not similar, knowledge should not be transferred. In such extreme situation, transfer learning can be unsuccessful and can lead to a \textit{negative transfer}. Prior work on transfer learning focused on ``\textit{What to transfer}'' and ``\textit{How to transfer}'', by implicitly assuming that the source and target are
related to each other. Hence, that work failed to address ``\textit{When to transfer}''~\cite{pan2010survey}. Jamshidi et al.~\cite{jamshidi2017transfer2}  alluded to this and explained when transfer learning works but, did not provide a method which can help in selecting a suitable source. In this paper, we focus on solving the problem of performance optimization by choosing a suitable source to transfer knowledge. This has not been explored in the prior work. 

The issue of identifying a suitable source is a common problem in transfer learning. To address this,
some researchers~\cite{krishna16,mensah17a, mensah17b, krishna18} have recently proposed the use of the \textit{bellwether} effect, which states that:
\begin{quote}
    `` ... When analyzing a community of software projects, then within that community there exists at least one exemplary project, called the \underline{bellwether(s)}, which can best define  predictors for other projects ...''
\end{quote}
The \textit{bellwether effect} has shown promise in identifying suitable sources for transfer learning in such varied domains as defect prediction, effort estimation, and code smell detection~\cite{krishna18}. However, the effect was shown only to work on domains where the data is relatively generic, easy to gather, and free of budget constraints~\cite{krishna18}. In this paper, we introduce BEETLE, a bellwether based transfer learner, which uses the bellwether effect to identify suitable sources to train transfer learners for performance optimization. 
Our main claim of this paper is:
\begin{quote}
A \textit{good source} with a \textit{simple transfer learner} is better than \underline{source agnostic} \textit{complex} transfer learners.
\end{quote}
In summary, we make the following contributions: 
\begin{enumerate}[leftmargin=*]
\item \textit{Source selection using bellwethers:} We show that the \textit{bellwether effect} exists in performance optimization and that we can use this to discover suitable sources (called bellwether environments) to perform transfer learning. (\S\ref{subsec:rq1})
\item \textit{A fast novel source selection algorithm:} We develop a fast algorithm for discovering the bellwether environment with only $\approx10\%$ of the measurements (\S\ref{sect:ourmethod}).
\item \textit{Transfer learning using Bellwethers:} We develop a novel transfer learning algorithm using Bellwether called BEETLE (short for  \underline{Be}llw\underline{e}ther \underline{T}ransfer \underline{Le}arner)  that uses the bellwether environment to construct a simple transfer learner (\S\ref{sect:ourmethod}).
\item \textit{More effective than non-transfer learning: } We show that using the BEETLE is just as good as than non-transfer learning approaches. It is also significantly more economical. (\~\S\ref{sect:rq2}).
\item \textit{More effective than state-of-the-art methods: } We show that the configurations discovered using the bellwether environment are much closer to the true-optima when compared to other state-of-the-art methods~\cite{valov2017transferring, jamshidi2017transfer}. And we show that we are a lot more economical. (~\S\ref{subsec:rq4}).

\end{enumerate}

The rest of the paper is structured as follows. In \tion{motivate}, provides motivations for this work. \tion{research_questions} introduces the research questions. In \tion{formalization}, we provide formal definitions of terminologies used in this paper. \tion{ourmethod} presents BEETLE, a new transfer learner proposed in this paper. In \tion{tl}, two state-of-the-art transfer learners are discussed. \tion{expt} contains the experimental setup. In \tion{results}, the results of the paper are presented as answers to  research questions. We discuss further implications of our findings in \tion{disc}. Threats of validity is highlighted in \tion{threats}. Finally, we conclude our findings in \tion{conclusion}.

\vspace{-0.6cm}
\section{Motivation}
\label{sect:motivate}
This section motivates our work by highlighting the many problems associated with performance modeling and transfer learning.

Modern software systems come with a large number of configuration options. 
For example, in {\sc Apache} (a popular web server) there are around 600 different configuration options and in {\sc hadoop}, as of version 2.0.0, there are around 150 different configuration options~\cite{xu2015hey}. Previous empirical studies have also shown that the number of options is growing over releases~\cite{xu2015hey}. These configuration options control the internal properties of the system such as memory, response times. The number of configuration options usually increase over time~\cite{van2017automatic, xu2015hey}. Given the \textbf{large number of configurations, it becomes increasingly difficult to assess the impact of the configuration options} of the performance of the software system. To address this issue, a common practice is to employ performance prediction models constructed using machine learning algorithms to estimate the performance of the system under these configurations~\cite{hoste2006, guo2013variability, hutter2014, thereska2010, valov2015, westermann12}. 

Further, research has shown that, when faced with a \textbf{large volume of configuration options, developers tend to ignore a majority (over 80\%) of the configuration options}~\cite{xu2015hey}. This leaves a considerable amount of untapped potential and often induces poor performance of software systems~\cite{xu2015hey}. To leverage the full benefit of the software system by exploiting the flexibility of the features offered by the system, researchers augment performance prediction models to enable performance optimization~\cite{nair2017using,oh2017finding}. Performance optimization extends performance prediction by identifying the best set of configuration options to pick to accomplish a given task with near-optimal performance. 


Performance optimization requires access to measurements of the software system under various configuration settings. However, obtaining these \textbf{performance measurements can take a significant amount of time and cost}. For example, in one of the software systems studied here (a video encoding application called {\sc x264}), it takes over 1536 hours to obtain performance measurements for 11 out the 16 possible configuration options~\cite{valov2017transferring}. This is in addition to other time-consuming tasks involved in commissioning these systems such as setup, teardown, etc. Further, making performance measurements can cost an exorbitant amount of money. For the same system, \fig{cost_x264} shows the amount we spent on gathering performance measurements on 2048 different configurations. 

For a software system under a new environment, instead of having to make exhaustive cost and time intensive measurements, it makes sense to \textbf{reuse performance measurements made for previous environments}. The concept of reusing information from other sources is the idea behind \textit{transfer learning}~\cite{Nam2013,Nam2015,Jing2015,kocaguneli2011find,kocaguneli2012,turhan09,peters15}. Specifically, to predict for the optimum configurations in a new environment (referred to as \textit{target environment}), we may use the performance measures of another workload as a proxy (referred to as the \textit{source environment}). For performance optimization, such transfer learning approaches have been shown to decrease the cost of learning by a significant amount~\cite{chen2011experience, jamshidi2017transfer, jamshidi2017transfer2, valov2017transferring}. 

\input{tex/cost_x264.tex}

It must be noted that \textbf{transfer learning methods place an implicit faith in the nature of the source}. Several researchers in transfer learning caution that the source must be chosen with care to ensure optimum performance of transfer learners~\cite{yosinski2014transferable, long2015, afridi2018}. An incorrect choice of source may result in the all too common \textit{negative transfer} phenomenon~\cite{ben2003, rosenstein2005,pan2010, afridi2018}. A negative transfer can be particularly damaging in that it often leads to performance degradation instead of performance optimization~\cite{jamshidi2017transfer2, afridi2018}. A preferred way to avoid negative transfer is with \textit{source selection}~\cite{afridi2018, krishna16, krishna18}. In software engineering, researchers have shown that the so-called \textit{bellwether effect} can be used to identify source datasets for effective transfer learning~\cite{krishna16}. This bellwether effect has been shown to be very effective in 
defect prediction, effort estimation, code-smell detection, etc.~\cite{mensah17a, mensah17b, krishna18}.

In this work, we introduce the notion of source selection with bellwether effect for transfer learning in performance optimization. With this, we develop a Bellwether Transfer Learner called BEETLE. We show that, for performance optimization, BEETLE can outperform both non-transfer learning methods and the current state-of-the-art transfer learning methods.

\section{Research questions}
\label{sect:research_questions}
This inquiry is structured around the following research questions.

\begin{itemize}[leftmargin=-1pt]
    \item[] \textbf{RQ1: Does there exist a Bellwether Environment?}\\
    \textit{\underline{\textbf{Purpose:}}} In the first research question, we ask if there exist bellwether environments to train transfer learners for performance optimization. We hypothesize that, if these bellwether environments exist, we can improve the efficacy of transfer learning algorithms. \\
    \textit{\underline{\textbf{Approach:}}} To answer this research question, we explore five popular open source software systems (for details see \tion{datasets} and ~\tab{datasets}). These software systems have performance measurements under different environments. In each of these software systems, we train a transfer learning model on one environment (the source) and predict for optima in all the other environments. We repeat this process in a round-robin manner for every environment. We then statistically rank the source environments based on their ability to find near-optimal solutions on the targets. If there exist bellwether environments, then those bellwether environments will have a better rank compared to all others environments.

    \vspace{0.15cm}
    \noindent\begin{minipage}{\linewidth}
    \begin{center}
    \arrayrulecolor{lightgray}
        \begin{tabular}{|p{0.95\linewidth}|}
            \hline
            \rowcolor{lightergray}\textit{\underline{\textbf{Result:}}} We find that bellwether environments are prevalent in performance optimization. That is, in each of the software systems, there exists at least one environment that can be used to construct superior transfer learners.\\\hline
        \end{tabular}
    \end{center}
    \end{minipage}
    \vspace{0.1cm}
    
    \item[] \textbf{RQ2: How many performance measurements are required to discover bellwether environments?}\\
    \textit{\underline{\textbf{Purpose:}}} Having established that bellwether environments are prevalent, the purpose of this research question is to establish how many performance measurements need to be made in the environments to discover bellwether environments.\\
    \textit{\underline{\textbf{Approach:}}} To answer this question, we developed an iterative method, based on incremental sampling strategy to find the bellwether environment. We start with 1\% of configurations from each environment and incrementally increases the number of sampled until we find the bellwether. Before each increment, we eliminate those environments which do not show much promise.\\[-0.1cm]
    
    \noindent\begin{minipage}{\linewidth}
    \begin{center}
        \arrayrulecolor{lightgray}
        \begin{tabular}{|p{0.95\linewidth}|}
            \hline
            \rowcolor{lightergray}\textit{\underline{\textbf{Result:}}} We can discover a potential bellwether environment by measuring as little as 10\% of the total configurations across all the software system.\\\hline
        \end{tabular}
    \end{center}
    \end{minipage}
    \vspace{0.1cm}

    \item[] \textbf{RQ3: How does BEETLE compare to non-transfer learning methods?}\\
    \textit{\underline{\textbf{Purpose:}}} The alternative to transfer learning is just to use the target data (similar to methods proposed in prior work) to find the near-optimal configurations. Literature is abundant with performance optimization algorithms that do not use transfer learning~\cite{guo2013variability, sarkar2015cost, nair2017using, nair2017faster}. For our comparisons, we used the performance optimization model proposed by Nair et al.~\cite{nair2017faster} in FSE '17. \\
    \textit{\underline{\textbf{Approach:}}} To answer this research question we compute the Win-Loss ratios of transfer learning with the bellwether environment (aka. BEETLE) to a regular performance optimization method. In addition to this, we compare the cost of the methods, in terms of number of measurements of learning a model. 
    
    \vspace{0.1cm}
    \noindent\begin{minipage}{\linewidth}
    \begin{center}
        \arrayrulecolor{lightgray}
        \begin{tabular}{|p{0.95\linewidth}|}
            \hline
            \rowcolor{lightergray}\textit{\underline{\textbf{Result:}}} Our experiments demonstrate that transfer learning using bellwethers (BEETLE) outperforms non-transfer learning methods both in terms of cost and the quality of the model.\\\hline
        \end{tabular}
    \end{center}
    \end{minipage}
    \vspace{0.1cm}
    
    \item[] \textbf{RQ4: How does BEETLE compare to state-of-the-art methods?}\\
    \textit{\underline{\textbf{Purpose:}}} In this research question we compare BEETLE with two other state-of-the-art transfer learners used commonly in performance optimization (for details see \tion{tl}). The purpose of this research question is to determine if a simple transfer learner like BEETLE with carefully selected source environments can perform as well as other complex transfer learning algorithms that do not perform source selection.\\
    \textit{\underline{\textbf{Approach:}}}
    To answer this question, we compare the performance of the near-optimal configurations found using the bellwether environment to the near-optimal configuration found by other transfer learning methods. The configuration found by the bellwether environment is similar to (or better than) the other transfer learning methods.\\[-0.2cm]
    
    \vspace{0.1cm}
    \noindent\begin{minipage}{\linewidth}
    \begin{center}
        \arrayrulecolor{lightgray}
        \begin{tabular}{|p{0.95\linewidth}|}
            \hline
            \rowcolor{lightergray}\textit{\underline{\textbf{Result:}}} We show that a simple transfer learning using bellwether environment (BEETLE) just as good as (or better than) current state-of-the-art transfer learners.\\\hline
        \end{tabular}
    \end{center}
    \end{minipage}
\end{itemize}

\section{Problem Formalization}
\label{sect:formalization}
\be[leftmargin=-1pt]
\item[] \textit{Environments:} A software system ($\mathcal{A}$) has $n$ configuration options which can be tweaked to change the performance of the software system ($Y$). The software system can be operated in different environments ($e\in E$). Each environment is described by 3 variables $\{h, w, v\}$ drawn from the environment space. Here, $h\in H$ represents the hardware, $w\in W$ represents the workload, and $v\in V$ represents the software version. In a software system, the total number of environments $E$ is given by $E=H\times W\times V$. A software system ($\mathcal{A}$) operating in an environment ($e\in E$) is denoted by $\mathcal{A}_e$.

\item[] \textit{Configuration:} Let $C_i$ indicate the $i^{th}$ configuration option of a software system $\mathcal{A}$ operating in environment $e$ (denoted by $\mathcal{A}_e$), which can either be (1)~numeric or (2)~boolean. A configuration $c^i$ is a member of the configuration space $\mathcal{C}$. $\mathcal{C}$ is a Cartesian product of all possible options $\mathcal{C}$ = Dom($C_1$) $\times$ Dom($C_2$) $\times$ ... $\times$ Dom($C_n$), where $\text{Dom}(C_i) = \{0, 1\}$ (in our setting) and $n$ is the number of configuration options. 

\item[]\textit{Performance:} Each configuration ($C$) of a system, $\mathcal{A}_e$, has a corresponding performance measure $y \in Y_{A,e}$ associated with it. The configuration and the corresponding performance measure is referred to as independent and dependent variables respectively. We denote the performance measure associated with a given configuration ($c^i$) by $y=f(c^i)$.  We consider the problem of finding the near-optimal configurations ($c^*$) such that $f(c^*)$ is less than other configurations in $C_{A,h,w,v}$. That is:
\begin{equation*}
\centering
     {\begin{array}{*{20}{l}}
     \centering
{f(c*) \le f(c){\rm{~~~~~~~~}}\forall c \in {C_{A,h,w,v}}\setminus c^* }&{{\text {for minimizing objective}}}\\
{f(c*) \ge f(c){\rm{~~~~~~~~}}\forall c \in {C_{A,h,w,v}}\setminus c^* }&{{\text {for maximizing objective}}}
\end{array}}
\end{equation*}
\item[]\textit{Transfer Learning:} In transfer learning, we find the near-optimal configuration for a target environment ($A_{e_t}$), by learning from the measurements ($<c,y>$) for the same system operating in different source environments ($A_{e_s}$). 
\item[]\textit{Bellwether Environment:} We show that, when performing transfer learning, there are exemplar source environments called the bellwether environment(s) ($\mathcal{B}=\{e_{s1}, e_{s2},...,e_{sn}\}\subset E$), which are the best source environment(s) to find near-optimal configuration for the rest of the environments ($\forall e \in E\setminus \mathcal{B}$). 
\ee

\section{{\large Beetle: \underline{BE}llw\underline{E}ther \underline{T}ransfer \underline{LE}arner}}
\label{sect:ourmethod}

\input{tex/approach.tex}

In this paper, we propose an alternative transfer learning approach to the current state-of-the-art discussed in the previous section. Our approach has two key components:
\be
    \item \textit{Identifying the bellwether environment:} To train a transfer model, we use bellwether effect to discover the best source environments (known as the \textit{bellwether environment}) among the available environments. 
    
    \item \textit{Construct the Transfer Model:} Next, to perform transfer learning, we use these bellwether environments to train a performance prediction model with regression tree~\cite{breiman1996bagging}.
\ee

Our key finding is that if a source environment is carefully selected using the bellwether effect, then it is possible to build a simple transfer model without any complex methods and still be able to generate near-optimal configurations in a target environment. 

In Figures~\ref{fig:approach_a} \&~\ref{fig:approach_b}, we describe BEETLE and list a generic algorithm of BEETLE. In this example, there are 7 source environments ($e_1, e_2,..., e_7$), which have been optimized previously. $e_8, e_9,..., e_{12}$ represents the target environments, which need to be optimized. BEETLE\textquotesingle objective is to find a bellwether among the source environments and use it to find the near-optimal configuration for the target environments. BEETLE, a Bellwether based approach can be separated into the following main steps: (i) finding the Bellwether environment, and (ii) using the Bellwether environment to find the near-optimal configuration for target environments. 

\subsection{Finding Bellwether Environments}
The central idea for finding Bellwether is to use minimal sampling to recursively eliminate environments which do not show promise of being a bellwether, that is those environments cannot be used to predict the near-optimal solutions. Figure~\ref{fig:approach_a} is a generic algorithm that defines the process of finding bellwethers. The process starts by sampling a small subset of the source environments. The size of the subset is controlled by a predefined parameter \textit{step\_size} (Line 6). The cost of sampling the configuration is calculated (Line 8). In our setting, we use the number of measurements as a proxy for cost and can be replaced by any user-defined cost function (\textit{get\_cost}). To compute the effectiveness of an environment, the sampling configurations along with the performance measure is used to build a performance model (regression tree in our setting). Please note that we choose regression tree as a model because it has been extensively used for performance prediction of configurable
software systems and demonstrated good results~\cite{guo2013variability, sarkar2015cost, nair2017faster, nair2017using, nair2018finding, guo2017data, valov2017transferring}. This model is then used to predict the optimal configuration among the configurations sampled in Line 6 (Line 10). We only used the sampled configurations because the actual performance of a source can only be calculated if the actual performance values (associated with the configurations) are known. This process is repeated for all the environments (represented as \textit{sources}) under consideration. Depending on how an environment can find the near-optimal configuration for other environments and a user-defined threshold (\textit{thres}), the non-bellwether environments are eliminated (Line 12). Non-bellwether environments are environments, which are not able to find near-optimal configurations for the other environments.  If the no environment is eliminated (with more
data) when compared to the previous iteration
(lesser data), then a life is lost (Line 14). When all lives are
expired or run out of the budget, the search process terminates. The environment with maximum performance is returned as the bellwether (Line 16). Please note that, \textit{FindBellwether} can identify multiple bellwethers ($e_1, e_2$). However, in our setting, we return a single bellwether.

The motivation behind using the parameter lives is to detect
convergence of the search process. If adding more configuration does
not improve the chances of finding the Bellwether, the search process should terminate to avoid resource wastage; see also \S~\ref{sect:disc}.

\vspace{-0.2cm}
\subsection{Using Bellwethers}

Once the bellwether is identified, it can be used to find the near-optimal configurations of target environments. As shown in Figure~\ref{fig:approach_b}   \textit{FindBellwether} returns the predicted bellwether environment (Line 3). Performance modeling then  samples the bellwether environments (Line 5)
for some number of samples (a user-defined parameter called budget). Please note that a user might choose to reuse the measurement used in \textit{FindBellwether} and save on cost. The sampled configuration and their corresponding performance measures it used to build a prediction model (Line 7). Similar to \textit{FindBellwether}, we use regression tree as our modeling method of choice. The prediction model is there used to predict the optimal configuration among the (unevaluated configurations) for the target environment (Lines 8-9). The predicted optimal configuration is returned as the best configuration. This process is then repeated for each target environments ($e_8, e_9,..., e_{12}$).

\vspace{-0.1cm}
\section{Transfer Learning in Performance Optimization}
\label{sect:tl}

In this paper,  BEETLE is compared against (a)~two state-of-the-art transfer learners from ICPC'17~\cite{van2017automatic} and SEAMS'17~\cite{jamshidi2017transfer}; and (b) a non-transfer learner from FSE'17~\cite{nair2017using}.

\vspace{-0.2cm}
\subsection{Transfer Learning with Linear Regression}

Valov et al.~\cite{valov2017transferring} proposed an approach for transferring performance models of software systems across platforms with \textit{different hardware settings}. \fig{lineartransform} shows the pseudocode of the transfer learning method. The method consists of the following two components: 
\bi
\item \textit{Performance prediction model:} The configuration source hardware are sampled using \textit{Sobol} sampling. The number of configurations is given by $T\times N_f$, where $T=\{3, 4, 5\}$ is the \textit{training coefficient} and $N_f$ is the number of configuration options. These configurations are used to construct a \textit{Regression Tree} model.
\item \textit{Transfer Model:} To transfer the predictions from the source to the target, the authors construct a linear regression model since it was found to provide good approximations of the transfer function. To construct this model, a small number of random configurations are obtained from the source and the target. 
\ei



\vspace{-0.2cm}
\subsection{Transfer Learning with Gaussian Process}

Jamshidi et al.~\cite{jamshidi2017transfer} took a slightly different approach to transfer learning. They used Gaussian Processes (GP) to find the relatedness between the performance measures in source and the target as well as the configurations. The relationships between input configurations were
captured in the GP model using a covariance matrix that
defined the kernel function to construct the Gaussian processes model. To encode the relationships between the measured performance of the configuration in the source the target, the authors propose a scaling factor to the above kernel. 

\input{tex/waterloo_peudo.tex}
\input{tex/pooyan_pseudo.tex}

\input{tex/datasets.tex}
The new kernel function is a defined as follows:
\begin{equation}
    k(s, t, f(s), f(t)) = k_t(s, t) \times k_{xx}(f(s), f(t))
\end{equation}\label{eq:gpkernel}
where $k_t(s,t)$ represents the multiplicative scaling factor. $k_t(s,t)$ is given by the correlation between source f(s) and target f(t) function, while $k_{xx}$ is the covariance function for input environments (s \& t). The essence of this method is that the kernel captures the interdependence  between the source and target configurations and their corresponding performance measurement values. 
\vspace{-0.2cm}
\subsection{Non-Transfer Performance Optimization}
\label{sect:nair}
A performance optimization model with no transfer was proposed by Nair et al.~\cite{nair2017using} in FSE '17. It works as follows: 
\be[leftmargin=*]
\item Sample a small set of measurements of configurations from the target environment
\item Construct performance prediction model with regression trees. 
\item Predict for near-optimal configurations. 
\ee
The key distinction here is that unlike transfer learners, that use a \textit{different source environment} to build to predict for near-optimal configurations in a target environment, a non-transfer method such as this uses configurations \textit{from within the target} environment to predict for near-optimal configurations.



\section{Experimental Setup}
\label{sect:expt}
\subsection{Subject Systems}
\label{sect:datasets}
In this study, we selected five configurable software systems from different domains, with different functionalities, and written in different programming. We selected these real-world software systems since their characteristics cover a broad spectrum of scenarios. \tab{datasets} lists the details of the software systems used here. The rest of this section provides a summary of the subject systems.

{\sc Spear} is an industrial strength bit-vector arithmetic decision procedure and Boolean satisfiability (SAT) solver. It is designed for proving software verification conditions, and it is used for bug hunting. We considered a configuration space with 14 options with $2^{14}$ or 16384 configurations. We measured how long it takes to solve an SAT problem in all 16,384 configurations in 10 environments.

{\sc x264} is a video encoder that compresses video files and has 16 configurations options to adjust output quality, encoder types, and encoding heuristics. Due to the size of the configuration space, we randomly sample 4000 configurations  in 21 environments. 

{\sc SQLite} is a lightweight relational database management system, embedded in several browsers and operating systems, which has 14 configuration options to change indexing and features for size compression. Due to a limited budget, we use 1000 randomly selected configurations in 15 different environments.

{\sc SaC} is a compiler for high-performance computing. The SaC compiler implements a large number of high-level and low-level optimizations to tune programs for efficient parallel executions. It has 50 configuration options to control optimization options. We measure the execution time of a program compiled in 71,267 randomly selected configurations. 

{\sc Storm} is a distributed stream processing framework which is used for data analytics. We run three benchmarks and measure the latency of the benchmark in 2,048 randomly selected configurations to assess the performance impact of Storm\textquotesingle s options.

\subsection{Evaluation Criterion}
\label{sect:eval}
Typically, performance models are evaluated based on the accuracy or error using measures such as MMRE. We note that there has been a lot of criticism regarding MMRE, which shows that MMRE along with other accuracy statistics such as MBRE has been shown to cause conclusion instability~\cite{myrtveit2012validity, myrtveit2005reliability, foss2003simulation}, given by:
\[
    MMRE = \frac{|predicted-actual|}{actual} \cdot 100
\]
While this typical in performance prediction, our objective is to find the near-optimal configurations or \textit{performance optimization}. For this, measures similar to MMRE is not applicable~\cite{nair2017using}. Recently work by Nair et al.~\cite{nair2017using}has shown that when MMRE cannot be used, other measures such as rank difference may be used---which emphasizes the sorted order of configurations and their performances rather than accuracy of the predictions.   

\input{RQ1.tex}

Once the performance model is trained, the accuracy of the model is measured  by sorting the values of $y=f(x)$ from `small' to `large', that is:
\begin{equation}
    f(c_1) \le f(c_2) \le f(c_3) \le ... \le f(c_n).
\end{equation}

\noindent The predicted rank order is then compared to the actual rank order.
We note that rank difference though effective is not particularly informative since it is sensitive to the workload. This means that in some workloads, a small difference in performance measure can lead to a large rank difference and vice-versa. For example, in {\sc Spear\_0} the top performance of the optimal configuration (rank 1) and 100$^{th}$ best configuration has a difference of 0.09\%---which means a large rank difference of 100 does not mean poor performance. 

Hence, we define a performance measure called \textit{Normalized Absolute Residual} (NAR), which represents the difference between the actual performance measurements of the optimal configuration and the predicted optimal configuration. The difference between the actual and predicted optimal configuration is normalized to the difference between the actual best and worst configurations, 
\begin{equation}
    \mathit{NAR} = \frac{min(f(c)) - f(c^{*})}{max(f(c)) - min(f(c))} \cdot 100  
\end{equation}\label{eq:nar}
where $c \in C_{A,h,w,v} \setminus c^*$. This measure is similar to Absolute Residual (lower is better). However, in our setting the range of the performance measures across different environments are not equal (hence the need for a normalization step). 


\subsection{Statistical Validation}
\label{sect:stats}

Our experiments discussed in RQ1 and RQ4 are all subject to inherent randomness introduced by sampling configurations or by different source and target environments. To overcome this, we use 30 repeated runs, each time with a different random number seed. The repeated runs provide us with a sufficiently large sample size for statistical comparisons. Each repeated run collects the values of NAR to assess the the transfer learners.

To rank these 30 numbers collected as above, we use the Scott-Knott test 
recommended by Mittas and Angelis~\cite{mittas13}. The Skott-Knott test has been endorsed by several SE researchers~\cite{leech2002call, poulding10, arcuri11, shepperd12a, kampenes07, Kocaguneli2013:ep}. Scott-Knott is a non-parametric statistical test that performs a bootstrap test with 95\% confidence~\cite{efron93} to determine the existence of statistically significant differences. This followed by an A12 test to check that any observed differences were not trivially small effects~\cite{Vargha00}. We say that a ``small'' effect has $a <0.6$. 
Scott-Knott test results in treatments being ranked from best to worst. Note that, if a set of treatments are not significantly different, they will have the same ranks.

\section{Results}
\label{sect:results}
\subsection*{RQ1: Does there exist a Bellwether Environment?}\label{subsec:rq1}

\textbf{\textit{\underline{Purpose:}}} The first research question seeks to establish the presence of  bellwether environments within different environments of a software system. We hypothesize that, if these bellwether environments exist, we can improve the  transfer learning algorithms.\\
\textbf{\textit{\underline{Approach:}}} For each subject software system, we use the environments to perform a pair-wise comparison method (similar to leave one out testing) as follows:
\be[leftmargin=*]
\item We pick one environment as a source and construct a transfer learner.
\item Next, we use the remaining environments as targets. For every target environment, we use the transfer learner-constructed in the previous step to predict for the optimum configuration.
\item Then, we measure the NAR of the predictions (see \tion{eval}). 
\item Afterward, we repeat steps 1, 2, and 3 for all the source environments and gather the outcomes.
\item Finally, we use Scott-Knott test to rank each environment (and its usefulness as a source).
\ee
\textbf{\textit{\underline{Summary:}}}
Our results are shown in Figure~\ref{fig:rq1}. Overall, we find that there is always at least one environment (the bellwether environment) in all the subject systems, that is much superior to others. Note that, {\sc Storm} is an interesting case, here all the environments are ranked 1, which means that all the environments are equally useful as a bellwether environment. Further, we note that the variance in the bellwether environments are much lower compared to other environments. 

\vskip 1ex
 \begin{myshadowbox}
         \textbf{\textit{\underline{Result:}}} There exist environments in each subject system, which act as bellwether environment and hence can be used to find the near-optimal configuration for the rest of the environments.
 \end{myshadowbox}

\begin{table}[t]
{\small
    \centering
    \caption{{\small Effectiveness of source selection method.}}
    \label{tbl:method}
    \begin{tabular}{@{}ccccc@{}}
    \toprule
    \multirow{2}{*}{\textbf{Subject System}} & \multicolumn{2}{c}{\textbf{\begin{tabular}[c]{@{}c@{}}Bellwether\\ Environment\end{tabular}}} & \multicolumn{2}{c}{\textbf{\begin{tabular}[c]{@{}c@{}}Predicted Bellwether\\ Environment\end{tabular}}} \\ \cmidrule(l){2-5} 
     & \textbf{Median} & \textbf{IQR} & \textbf{Median} & \textbf{IQR} \\ \midrule
    SQLite & 0.8 & 1.13 & 1.8 & 2.48 \\
    Spear & 0.1 & 0.1 & 0.1 & 0 \\
    X264 & 0.35 & 1.62 & 0.9 & 1.06 \\
    Storm & 0.0 & 0.0 & 0.0 & 0.0 \\
    Sac & 0.27 & 0.14 & 0.63 & 7.4 \\ \bottomrule
    \end{tabular}
}
\end{table}

\vspace{-0.25cm}
\subsection*{RQ2: How many performance measurements are required to discover bellwether environments?}

\textbf{\textit{\underline{Purpose:}}} The bellwether environments found in RQ1 required us to use 100\% measurements from all the environments. This may not be economical in a real-world scenario. It can be prohibitively
expensive to run and test all configurations of subject systems (as done in RQ1) since their configuration spaces are large. Thus, in this research question, we ask if we can find the bellwether environments sooner using fewer configurations. \\
\textbf{\textit{\underline{Approach:}}}
We developed an iterative method, based on incremental sampling strategy to find the bellwether environment. It works as follows
\be
\item We start from 1\% of configurations from each environment and assume that every environment is a potential bellwether environment.
\item Then, we increment the number of configurations in steps of 1\% and measure the NAR values. 
\item We rank the environments and eliminate those that do not show much promise. A detailed description of how this is accomplished can be found in~\S\ref{sect:ourmethod}. 
\item We repeat the above steps until we cannot eliminate any more environments.
\ee

The above strategy uses $\mu+\sigma$ of the NAR values at each step as a threshold to eliminate non-bellwether environments. To function correctly, this requires the NAR values to follow a normal distribution. If normality is violated, we used power transforms to make the data more normal distribution-like. We note that this is a prevalent strategy commonly used in other domains to reduce the size of options or alternatives~\cite{borgelt2005keeping} and is also known as \textit{backward elimination}~\cite{blum1997selection}.

To see if our proposed method is effective, we compare the performance of bellwether environment with the predicted bellwether environment.\\
\textbf{\textit{\underline{Summary:}}}
Table~\ref{tbl:method} summarizes our findings. We find that, 
\begin{itemize}[leftmargin=*]
    \item In all 5 cases, using at most 10\% of the configurations we find one of the bellwether environments that are found with 100\% of the configurations. See, column Rank in Table~\ref{tbl:method}. 
    \item In terms of quality of predictions, the NAR values of the predicted bellwether environments with 10\% of the configurations is less than 1.0\% different from the bellwether found at 100\%.
\end{itemize}

Our results are encouraging in that they demonstrate how the bellwether environments can be discovered very fast with just a fraction of the original configuration size. Since fewer configuration takes less time to collect and is cheaper, we can assert that discovering bellwether environments can be very economical. 

 \vskip 1ex
 \begin{myshadowbox}
         \textbf{\textit{\underline{Result:}}} The bellwether environment can be recognized using only a fraction of the measurements (under 10\%), and the identified bellwether environments have similar NAR values to the actual bellwether environment. 
 \end{myshadowbox}

\vspace{-0.3cm}
\subsection*{RQ3: How does BEETLE compare to non-transfer learning methods?}
\label{sect:rq2}

\textbf{\textit{\underline{Motivation}}}: Having established that there exist bellwether environments in the subjects systems (RQ1) and that they can be found with very few measurements (RQ2), in this research question we explore how BEETLE  compares to a non-transfer learning approach. For our experiment, we use the non-transfer performance optimizer proposed by Nair et al.~\cite{nair2017using}. More details on Nair et al.'s method can be found in \tion{nair}.

\begin{figure}[!b]
    \centering
    \includegraphics[width=\linewidth]{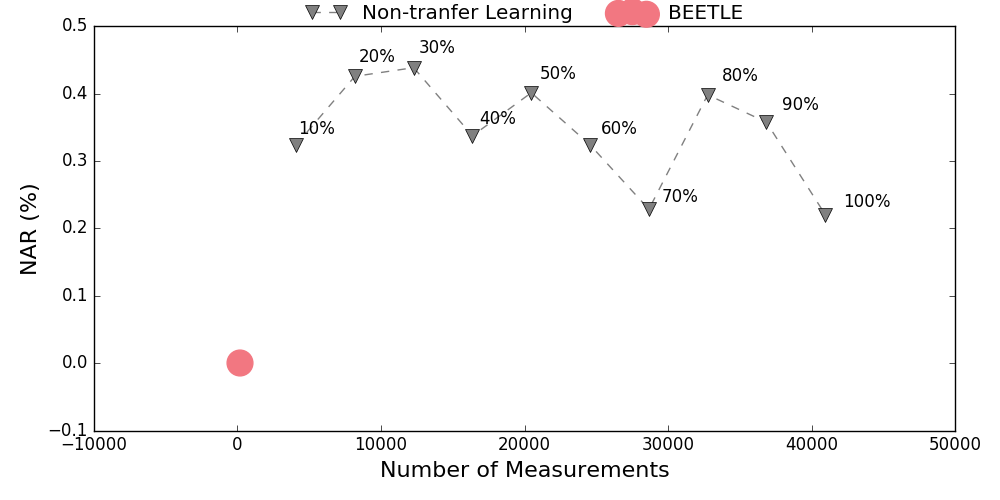}
    \caption{{\small Trade-off of between the quality of the configurations and the cost to build the model in case of {\sc X264}. We notice that the cost to find a good configuration using bellwethers is much lower than that of Non-transfer learning methods. }}
    \label{fig:tradeoffx264}
\end{figure}

\input{RQ2.tex}

Both BEETLE and Nair et al.'s methods seek to achieve the same goal---find optimal configuration in a target environment. BEETLE uses configurations from a \textit{different source} to achieve this, whereas the non-transfer learner uses configurations from \textit{with the target}.\\
\textbf{\textit{\underline{Approach:}}} Our setup involves evaluating the Win/Loss ratio of BEETLE to the non-transfer learning algorithm while predicting for the optimal configuration. Comparing against true optima, we define ``win'' as cases where BEETLE has a better (or same) optima as the non-transfer learner. A ``loss'' otherwise.\\
\textbf{\textit{\underline{Summary:}}}
Our results are shown in Figures~\ref{fig:tradeoffx264} and \ref{fig:rq2_1}. In \fig{rq2_1}, the x-axis represents the number of configurations (expressed in \%) to train the non-transfer learner and BEETLE, and the y-axis represents the number of wins/losses. From this figure we observe:
\begin{itemize}[leftmargin=*]
\item \textit{Better performance:} In 4 out of 5 systems, the BEETLE ``wins'' significantly more than it ``losses''. This means that BEETLE is better than (or at least as good as) non-transfer learning methods. 
\item \textit{Lower cost:} In terms of cost, we note that BEETLE outperforms the non-transfer learner significantly, ``winning'' at configurations of 10\%  to 100\% of the original sample size. Further, when we look at the trade-off between performance and number of measurements in \fig{tradeoffx264}, we note that BEETLE achieves an NAR close to zero with close around 100 samples. On the other hand, the non-transfer learning method of Nair et al.~\cite{nair2017using}has significantly larger NAR while also requiring large sample sizes.\\
\end{itemize}
\vskip 1ex
\begin{myshadowbox}
     \textbf{\textit{\underline{Result:}}} BEETLE performs better than (or same as) a non-transfer learning approach. BEETLE is also cost/time efficient as it requires far fewer measurements.
\end{myshadowbox}

\vspace{-0.1cm}
\subsection*{RQ4: How does BEETLE compare to state-of-the-art methods?}\label{subsec:rq4}
\input{RQ4.tex}

\noindent\textbf{\textit{\underline{Purpose:}}} The main motivation of this work is to show that the source environment can have a significant impact on transfer learning.  In this research question, we seek to compare BEETLE with two other state-of-the-art transfer learners by Jamshedi et al.~\cite{jamshidi2017transfer} and Valov et al.~\cite{valov2017transferring}. For further details of these methods, see \tion{tl}.\\
\textbf{\textit{\underline{Approach:}}} 
To perform our comparisons, we use a  Scott-Knott test to rank the $NAR$ values. These $NAR$ values indicate the \% performance difference between estimated and the actual near-optimal.\\
\textbf{\textit{\underline{Summary:}}}
Our results are shown in Figure~\ref{fig:rq4}. In this figure, the best transfer learner is ranked 1. We note that in 4 out of 5 cases, the baseline transfer learner based on source selection performs just as well as (or better than) the state-of-the-art. This result is encouraging in that it points to significant impact choosing a good source environment can have on the performance of transfer learners. Further, in ~\fig{rq4_measurement} we compare the number of performance measurements required to construct the transfer learners (note the logarithmic scale on the vertical axis). It can be noticed that BEETLE requires far fewer measurements compared to the other transfer-learning methods.

\vskip 1ex
 \begin{myshadowbox}
         \textbf{\textit{\underline{Result:}}} In most software systems, BEETLE performs just as well as (or better than) other state-of-the-art transfer learners for performance optimization using far fewer measurements.
 \end{myshadowbox}

\section{Discussion}
\label{sect:disc}



\begin{figure}[t]
    \centering
    \includegraphics[width=0.98\linewidth, height=4cm]{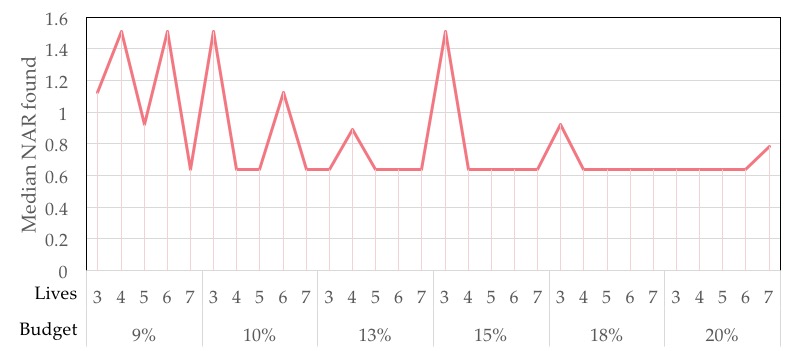}
    \caption{The trade-off between the budget of the search, the number of lives, and the NAR (quality) of the solutions.}
    \label{fig:tuning}
\end{figure}

\noindent \textbf{What is the trade-off between hyper-parameters and effectiveness of BEETLE?}
In Figure~\ref{fig:tuning}, we show the trade-off between the hyperparameters (budget, lives) and NAR values (effectiveness). We note that the performance is correlated to the budget and number of lives. That is, as budget increases the NAR value decreases. Since our objective is to minimize the number of measurements while reducing overall NAR, we assign the value of 5 to lives and 10\% to budget for our experiments.




\noindent\textbf{When are bellwethers ineffective?}
Existence or discovery of bellwethers depends on the following: (a)~\textit{Metrics used:} Finding bellwether using metrics that are not justifiable, may be unsuccessful, e.g, trying to discover bellwethers in performance optimization, by measuring MMRE instead of NAR for will fail (see \url{http://tiny.cc/bw_metrics})~\cite{nair2017using}; (b)~\textit{Different Software System:} Bellwethers of a certain software system `A' may not work for software system `B'; and (c) \textit{Different Performance Measures: } Bellwether discovered for one performance measure (time) may not work for other performance measures (throughput).

\noindent\textbf{Is BEETLE applicable in other domains?} Yes, BEETLE can be applied to any transfer learning application, where the choice of the source data impacts the performance of transfer learning. This can be applied to problems such as configuring big data systems~\cite{JC:MASCOTS16}, finding suitable cloud configuration for a workload~\cite{Hsu2018scout, hsu2017low}, 
configuring hyper parameters of machine learning algorithms~\cite{fu2016tuning, fu2016differential, afridi2018}. 

\section{Threats to Validity}
\label{sect:threats}

\noindent{\em External validity:} We selected a diverse set of subject systems and a large number of selected environment changes, but, as usual, one has to be careful when generalizing to other subject systems and environment changes. Even though we tried to run our experiment on a variety of software systems from different domains, we cannot generalize our results beyond these software systems.


\noindent{\em Internal validity}: Due to the size of configuration spaces, we could only measure configurations exhaustively in one subject system and had to rely on sampling (with substantial sampling size) for the others, which may miss effects in parts of the configuration space that we did not sample. We did not encounter any surprisingly different observation in our exhaustively measured {\sc SPEAR} dataset. 
Measurement noise in benchmarks can be reduced but not avoided. We performed benchmarks on dedicated systems and repeated each measurement 3 times. We repeated experiments when we encountered unusually large deviations.

\noindent{\em Parameter bias: }With all the transfer learners and predictors discussed here, there
are a number of internal parameters that have been set by default.
The result of changing these parameters may (or may not) have
a significant impact on the outcomes of this study.

\section{Conclusion}
\label{sect:conclusion}
Our approach exploits the bellwether effect---there are one or more bellwether environments which can be used to find good configurations for rest of the environments. We also propose a new transfer learning method, called BEETLE, which exploits this phenomenon. We show that BEETLE can quickly identify the bellwether environments with only a few measurements ($\approx10\%$) and use it to find the near-optimal solutions in the target environments. We have done extensive experiments with 5 highly configurable systems demonstrating that BEETLE can (i)~identify the most suitable source to construct transfer learners, (ii)~find near-optimal configurations with only a small number of measurements (less than $13.5\%\approx1/7^{th}$ of configuration space), (iii)~performs as well as non-transfer learning approaches
, and (iv)~performs as well as state-of-the-art transfer learners.

\bibliographystyle{plain}
\balance
\input{cells.bbl}

\end{document}

%% file: tex/cost_x264.tex
\begin{figure}[t!]
    \centering
    \includegraphics[width=.9\linewidth]{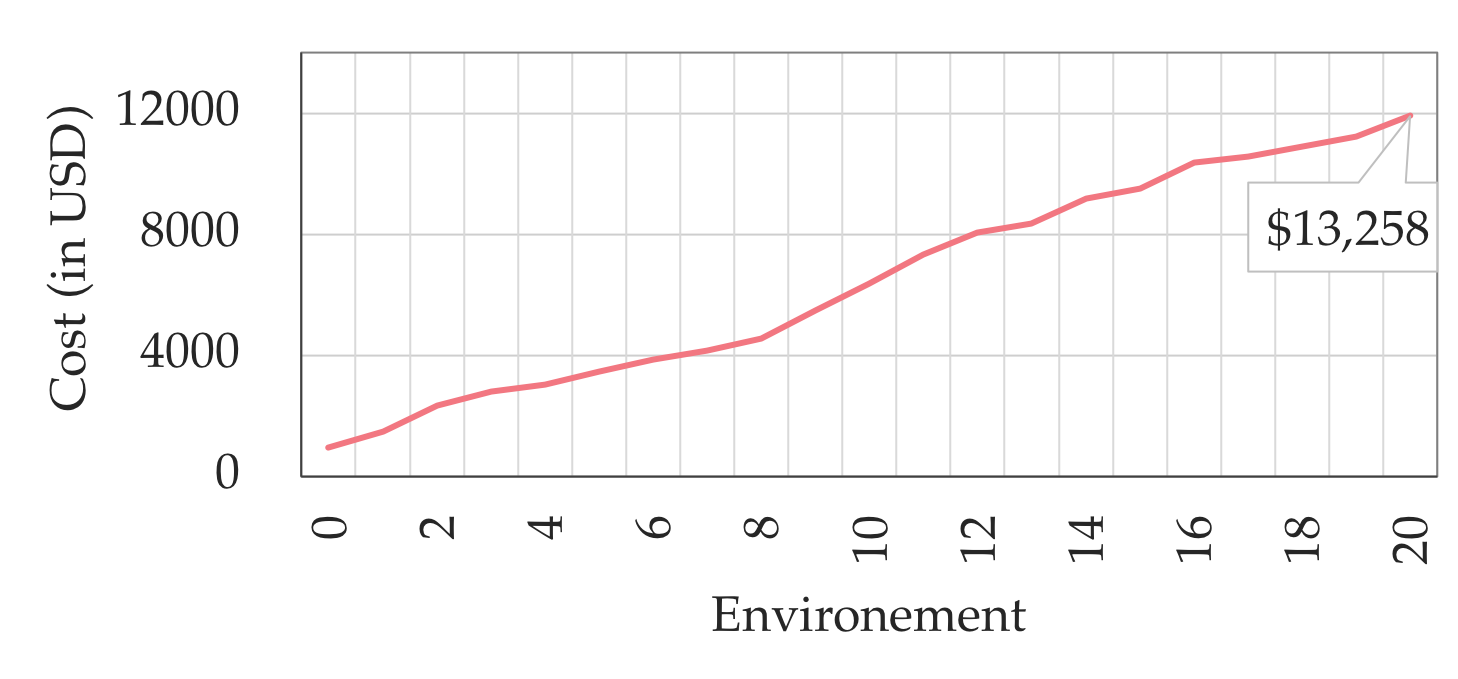}
    \caption{{\small Cost incurred for performance measurements of 2048 configuration options for {\sc x264}. Measurements were made on Amazon AWS \texttt{c4.large} cluster at \$0.0252/hour.}}
    \label{fig:cost_x264}
\end{figure}

%% file: tex/approach.tex
\begin{figure}[t]
\centering
\begin{subfigure}[t]{\linewidth}
\centering
\textbf{Pick Bellwether Environment}\\[0.1cm]
\includegraphics[width=\linewidth]{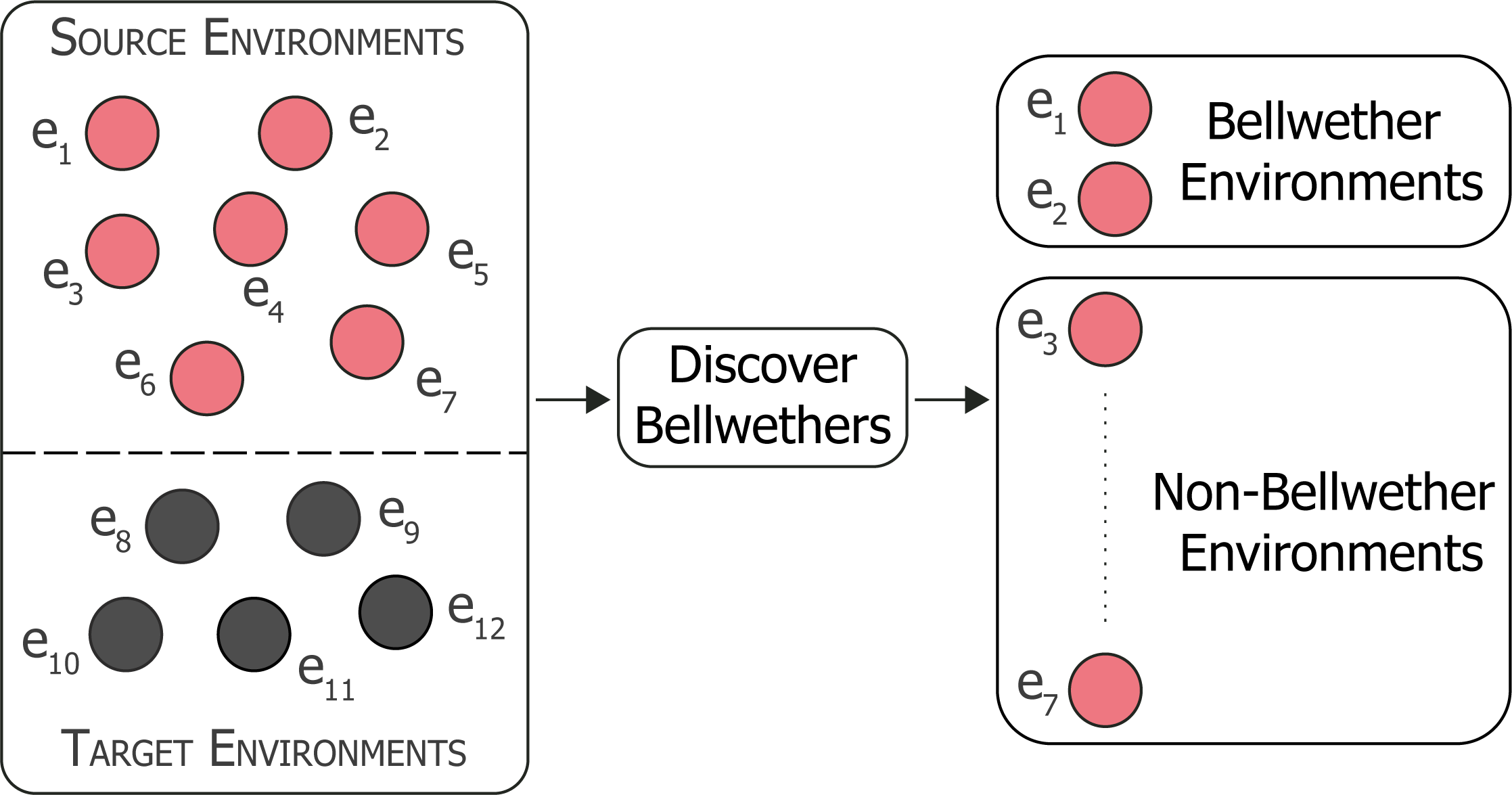}
\end{subfigure}\\
\begin{subfigure}[t]{\linewidth}
\vspace{0.5cm}
\small
\begin{lstlisting}[xleftmargin=4.0ex,mathescape,frame=none,numbers=left]
def FindBellwether(sources, step_size, budget, thres, lives): 
  while lives or cost > budget:
   "Sample configurations"
   sampled = list()
   for source in sources:
     sampled += source.sample(step_size)
   "Get cost"
   cost = get_cost(sampled)
   "Evaluate pair-wise performances"
   perf = get_perf(sampled)
   "Remove non-bellwether environments"
   sources=remove_non_bellwethers(sources, perf, thres)
   "Loose life if no sources are removed"
   if prev == len(sources): lives -= 1
   "Return a bellwether"
  return sources[argmin(perf)]
\end{lstlisting}
\end{subfigure}
\caption{{\small This figure demonstrates how to pick the bellwethers}}
\label{fig:approach_a}
\end{figure}

\begin{figure}[t]
\begin{subfigure}[t]{\linewidth}
\centering
\textbf{Transfer Learning with Bellwether Environment}\\
\includegraphics[width=0.85\linewidth]{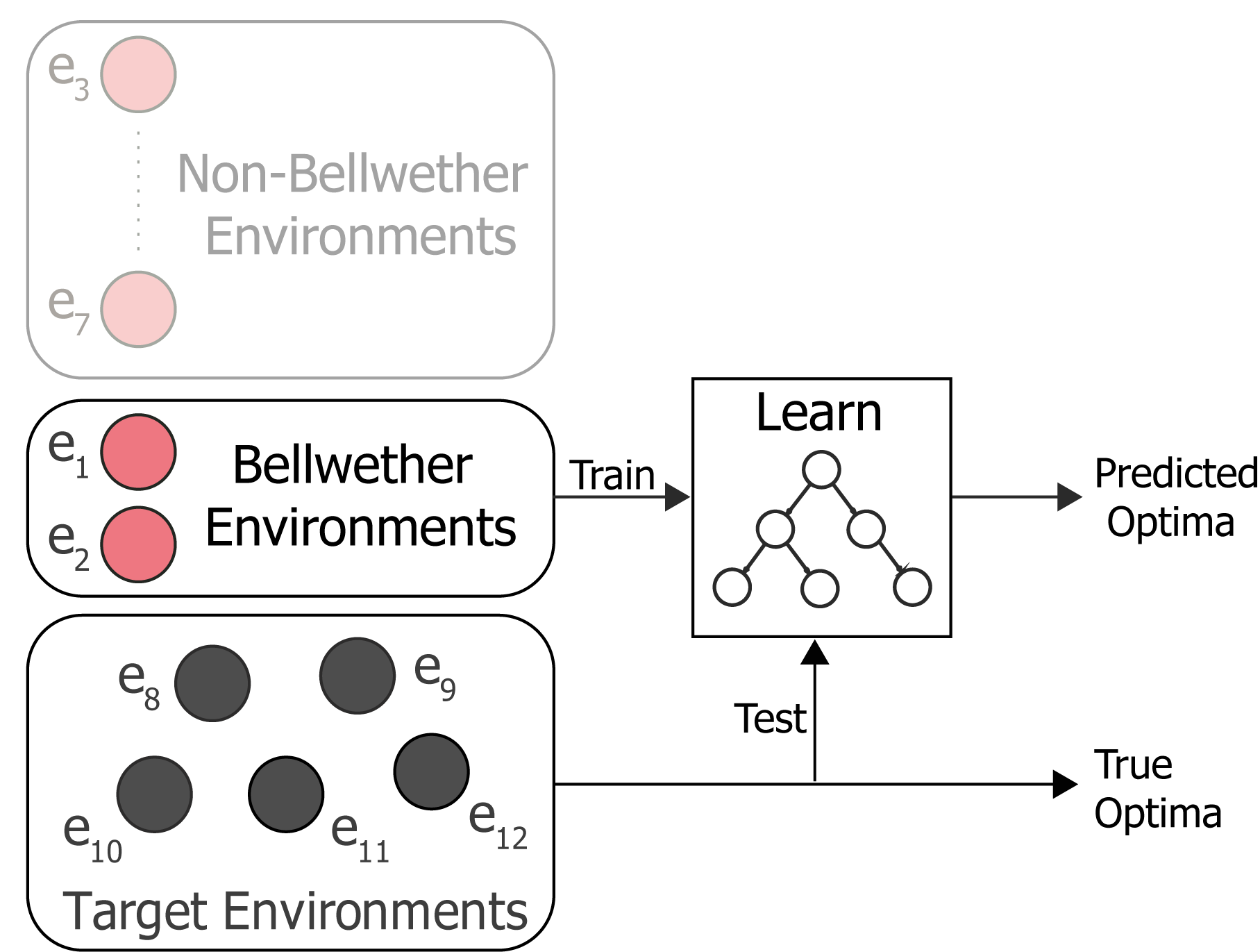}
\end{subfigure}~~\\
\begin{subfigure}[t]{\linewidth}
\vspace{0.4cm}
\small
\begin{lstlisting}[xleftmargin=5.0ex,mathescape,frame=none,numbers=left]
def BEETLE(sources, target, budget): 
  "Find the bellwether environment"
  bellwether = FindBellwether(sources, step_size, budget, thres, lives)
  "Sample the bellwether source to fit budget"
  b_some = bellwether.sample(budget)
  "Train a prediction model with the bellwether"
  prediction_model = regTree.train(b_some)
  predicted = prediction_model.test(target.indep)
  return target[argmin(predicted)]
\end{lstlisting}
\end{subfigure}	
\caption{{\small This figure demonstrates how to construct BEETLE.}}
\label{fig:approach_b}
\end{figure}

%% file: tex/waterloo_peudo.tex
\begin{figure}[t]
\small
\begin{lstlisting}[xrightmargin=6ex, mathescape,frame=none,numbers=right]
def LinearTransform(source, target, 
                    training_coef, budget): 
  "Construct a prediction model"
  prediction_model=regTree.train(source, training_coef)
  "Sample random measurements"
  s_samp = source.sample(budget)
  t_samp = target.sample(budget)
  "Get performance measurements"
  s_perf=get_perf(s_samp)
  t_perf=get_perf(t_samp)
  "Train a transfer model with LR"
  transfer_model=linear_model.train(s_perf, t_perf)
  return prediction_model, transfer_model
\end{lstlisting}
\caption{\small{Linear Transformation Transfer. From Valov et al.~\cite{valov2017transferring}.}}
\label{fig:lineartransform}  
\end{figure}

%% file: tex/pooyan_pseudo.tex
\begin{figure}[t]
\small
\begin{lstlisting}[xrightmargin=6ex, mathescape,frame=none,numbers=right]
def GPTransform(source, target, 
                 source_budget, target_budget): 
  "Sample random configurations"
  s_some = source.sample(source_budget)
  t_some = target.sample(target_budget)
  "Get performance measurements"
  s_perf = get_perf(s_some)
  t_perf=get_perf(t_some)
  "Compute correlation"
  perf_correlation = get_correlation(s_perf, t_perf)
  "Compute covariance"
  input_covariance = get_covariance(s_some, t_some)
  "Construct a kernel"
  kernel = input_covariance $\times$ perf_correlation
  "Train the Gaussian Process model"
  learner = GaussianProcessRegressor(kernel)
  prediction_model = learner.train(s_some)
  return prediction_model
\end{lstlisting}
\caption{\small{Gaussian Process Transformation Transfer. From   Jamshidi et al.~\cite{jamshidi2017transfer}.}}
\label{fig:gptransform}  
\end{figure}

%% file: tex/datasets.tex
\begin{table*}[t]
\caption{{\small Overview of the real-world subject systems. |C|=Number of Configurations, |c|=Number of configuration options, |E|: Number of Environments, |H|: Hardware, |W|: Workloads, and |V|: Versions. See \url{http://tiny.cc/bw_software} for more details.}}
{\small
    \centering
    \label{my-label}
    \begin{tabular}{@{}ccp{8cm}p{0.5cm}p{0.5cm}p{0.5cm}p{0.5cm}p{1cm}p{0.5cm}@{}}
    \toprule
    \multicolumn{1}{c}{\textbf{Domain}} & \multicolumn{1}{c}{\textbf{System}} & \multicolumn{1}{c}{\textbf{Description}} & \multicolumn{1}{c}{\textbf{|C|}} & \multicolumn{1}{c}{\textbf{|c|}} & \multicolumn{1}{c}{\textbf{|E|}} & \multicolumn{1}{c}{\textbf{|H|}} & \multicolumn{1}{c}{\textbf{|W|}} & \multicolumn{1}{c}{\textbf{|V|}} \bigstrut\\ \midrule
    \rowcolor[HTML]{EFEFEF} 
\begin{tabular}[c]{@{}l@{}}Video\bigstrut\\  Encoder\end{tabular} & X264                                                                                   & x264is a video encoder that compresses video files  to adjust output quality, encoder types,and encoding heuristics.                                                                                  & 4000                             & 16                               & 21                               & 4                                & 3                                & 3                                \bigstrut\\
\begin{tabular}[c]{@{}l@{}}SAT\bigstrut\\ Solver\end{tabular}     & SPEAR                                                                                  & An industrial strength bit-vector arithmetic decisionprocedure and a Boolean satisfiability (SAT) solver. It is designed for proving software verification conditions and it is used for bug hunting. & 16384                            & 14                               & 10                               & 2                                & 4                                & 2                                \bigstrut\\
\rowcolor[HTML]{EFEFEF} 
Database                                                 & SQLite                                                                                 & x264is a video encoder that compresses video files  to adjust output quality, encoder types,and encoding heuristics.                                                                                  & 1000                             & 14                               & 15                               & 2                                & 13                               & 2                                \bigstrut\\
Compiler                                                 & SaC                                                                                    & SQLite is a lightweight relational database management sys-tem, embedded in several browsers and operating systems.                                                                                   & 846                              & 50                               & 7                                & 2                                & 10                               & 2                                \bigstrut\\
\rowcolor[HTML]{EFEFEF} 
\begin{tabular}[c]{@{}l@{}}Data\bigstrut\\ Analytics\end{tabular} & \begin{tabular}[c]{@{}l@{}}Apache\bigstrut\\ Storm\end{tabular}                                 & Apache Storm is a distributed stream processing computation framework written predominantly in the Clojure programming language.                                                                      & 2048                             & 12                               & 4                                & 2                                & 3                                & 2                                \bigstrut\\ \hline

    \end{tabular}%
    \label{tab:datasets}
}

\end{table*}

%% file: RQ1.tex
\begin{figure*}[t]
{\small
    \begin{minipage}[]{0.475\linewidth}
    \textbf{{\sc X264}}\\
    \resizebox{\linewidth}{!}{%
    
    \arrayrulecolor{lightgray}%
    \begin{tabular}{|llrrc|}
    \arrayrulecolor{lightgray}
    \rowcolor{lightgray}{\small \textbf{Rank}} & {\small\textbf{Dataset}} & {\small \textbf{Median}} & {\small\textbf{IQR}} & \\\hline  
    \rowcolor{lightergray}  1 &      x264\_18 &    0.35  &  1.82 & \quart{0}{2}{0}{1} \\
    \rowcolor{lightergray}  1 &       x264\_9 &    0.35  &  1.62 & \quart{0}{2}{0}{1} \\
    \hline  2 &      x264\_10 &    0.94  &  8.25 & \quart{0}{12}{1}{1} \\
      2 &       x264\_7 &    0.94  &  8.25 & \quart{0}{12}{1}{1} \\
      2 &      x264\_11 &    1.62  &  7.46 & \quart{1}{11}{2}{1} \\
     3 &      x264\_16 &    2.33  &  12.18 & \quart{0}{18}{3}{1} \\
      3 &       x264\_2 &    2.33  &  12.18 & \quart{0}{18}{3}{1} \\
    3 &       x264\_6 &    2.82  &  5.35 & \quart{0}{8}{4}{1} \\  
    3 &      x264\_20 &    3.65  &  13.74 & \quart{1}{20}{5}{1} \\
      4 &      x264\_19 &    6.95  &  41.97 & \quart{3}{62}{10}{1} \\
      4 &       x264\_3 &    8.68  &  49.78 & \quart{6}{73}{12}{1} \\
        4 &      x264\_17 &    13.61  &  32.32 & \quart{5}{48}{20}{1} \\
      4 &      x264\_13 &    16.42  &  51.65 & \quart{2}{76}{24}{1} \\
      4 &      x264\_15 &    20.14  &  50.68 & \quart{4}{75}{29}{1} \\
      5 &      x264\_14 &    27.24  &  42.74 & \quart{3}{63}{40}{1} \\
      5 &       x264\_0 &    28.63  &  49.77 & \quart{6}{73}{42}{1} \\
    \hline \end{tabular}}\\
    \textbf{{\sc Sac}}\\
    \resizebox{\linewidth}{!}{%
    \begin{tabular}{|llrrc|}
    \arrayrulecolor{lightgray}
    \rowcolor{lightgray}{\small \textbf{Rank}} & {\small\textbf{Dataset}} & {\small \textbf{Median}} & {\small\textbf{IQR}} & \\\hline
    \rowcolor{lightergray}  1 &        sac\_6 &    0.27  &  0.14 & \quart{0}{0}{0}{0} \\
    \hline  2 &        sac\_4 &    0.96  &  4.26 & \quart{0}{3}{0}{0} \\
      2 &        sac\_8 &    1.04  &  3.67 & \quart{0}{3}{0}{0} \\
      2 &        sac\_9 &    2.29  &  4.98 & \quart{1}{4}{1}{0} \\
      3 &        sac\_5 &    10.8  &  89.65 & \quart{8}{71}{8}{0} \\
    \hline \end{tabular}}\\
    \textbf{{\small {\sc Storm} }}\\
    \resizebox{\linewidth}{!}{%
    \begin{tabular}{|llrrc|}
    \arrayrulecolor{lightgray}
    \rowcolor{lightgray}{\small \textbf{Rank}} & {\small\textbf{Dataset}} & {\small \textbf{Median}} & {\small\textbf{IQR}} & \\\hline  
    \rowcolor{lightergray}  1 & storm\_feature9 &    0.0  &  0.0 & \quart{0}{0}{0}{1999} \\
    \rowcolor{lightergray}  1 & storm\_feature8 &    0.0  &  0.0 & \quart{0}{0}{0}{1999} \\
    \rowcolor{lightergray}  1 & storm\_feature6 &    0.0  &  0.01 & \quart{0}{19}{0}{1999} \\
    \rowcolor{lightergray}  1 & storm\_feature7 &    0.01  &  0.04 & \quart{0}{79}{19}{1999} \\
    \hline \end{tabular}}
    \end{minipage}\hspace{10pt}
    \begin{minipage}[]{0.475\linewidth}
    \textbf{{\small {\sc Spear}}}\\[0.1cm]
    \resizebox{\linewidth}{!}{%
    \begin{tabular}{|llrrc|}
    \arrayrulecolor{lightgray}
    \rowcolor{lightgray}{\small \textbf{Rank}} & {\small\textbf{Dataset}} & {\small \textbf{Median}} & {\small\textbf{IQR}} & \\\hline  
    \rowcolor{lightergray} 1 &      spear\_7 &    0.1  &  0.1 & \quart{0}{1}{1}{13} \\
    \rowcolor{lightergray}  1 &      spear\_6 &    0.1  &  0.2 & \quart{0}{2}{1}{13} \\
    \rowcolor{lightergray}  1 &      spear\_1 &    0.1  &  0.1 & \quart{0}{1}{1}{13} \\
    \rowcolor{lightergray}  1 &      spear\_9 &    0.1  &  0.5 & \quart{0}{6}{1}{13} \\
    \rowcolor{lightergray}  1 &      spear\_8 &    0.1  &  0.2 & \quart{0}{2}{1}{13} \\
    \rowcolor{lightergray}  1 &      spear\_0 &    0.1  &  0.91 & \quart{0}{12}{1}{13} \\
    \hline  2 &      spear\_5 &    0.28  &  0.3 & \quart{2}{4}{3}{13} \\
      3 &      spear\_4 &    0.6  &  1.17 & \quart{5}{16}{8}{13} \\
      4 &      spear\_2 &    1.09  &  5.31 & \quart{8}{71}{14}{13} \\
      5 &      spear\_3 &    1.89  &  4.48 & \quart{18}{61}{25}{13} \\
    \hline \end{tabular}}\\[0.2cm]
    \textbf{{\small {\sc Sqlite}}}\\[0.1cm]
    \resizebox{\linewidth}{!}{%
    \begin{tabular}{|llrrc|}
    \arrayrulecolor{lightgray}
    \rowcolor{lightgray}{\small \textbf{Rank}} & {\small\textbf{Dataset}} & {\small \textbf{Median}} & {\small\textbf{IQR}} & \\\hline  
    \rowcolor{lightergray}  1 &    sqlite\_17 &    0.8  &  1.13 & \quart{0}{1}{0}{0} \\
    \rowcolor{lightergray}  1 &    sqlite\_59 &    2.0  &  3.44 & \quart{0}{4}{2}{0} \\
    \rowcolor{lightergray}  1 &    sqlite\_19 &    2.0  &  4.88 & \quart{0}{6}{2}{0} \\
    \hline  2 &    sqlite\_44 &    1.96  &  6.91 & \quart{0}{10}{2}{0} \\
      2 &    sqlite\_16 &    2.52  &  7.41 & \quart{1}{10}{3}{0} \\
      2 &    sqlite\_73 &    2.82  &  7.24 & \quart{1}{10}{3}{0} \\
      2 &    sqlite\_45 &    3.47  &  11.86 & \quart{0}{17}{4}{0} \\
      2 &    sqlite\_10 &    3.88  &  6.92 & \quart{1}{9}{4}{0} \\
      2 &    sqlite\_96 &    4.94  &  6.04 & \quart{1}{8}{6}{0} \\
      2 &    sqlite\_79 &    5.64  &  5.24 & \quart{4}{7}{7}{0} \\
      2 &    sqlite\_11 &    6.64  &  5.75 & \quart{5}{8}{8}{0} \\
      2 &    sqlite\_52 &    6.84  &  7.95 & \quart{1}{11}{8}{0} \\
      2 &    sqlite\_97 &    7.68  &  13.71 & \quart{7}{18}{10}{0} \\
      3 &    sqlite\_18 &    13.17  &  54.68 & \quart{5}{74}{17}{0} \\
      3 &    sqlite\_94 &    27.43  &  47.66 & \quart{9}{65}{37}{0} \\
    \hline \end{tabular}}
    \end{minipage}
    \caption{ {\small Median NAR of 30 repeats. Median NAR is the normalized absolute residual values  as described in Equation~\ref{eq:nar}, and IQR the difference between 75th percentile and 25th percentile found during multiple repeats. Lines with a dot in the middle (~\protect\quartex{-2}{13}{6}), show the median as a round dot within the IQR. All the results are sorted by the median NAR: a lower median value is better. The left-hand column (\textit{Rank}) ranks the various techniques where lower ranks are better.}}
    \label{fig:rq1}
}
\end{figure*}

%% file: RQ2.tex
\begin{figure*}[t]
	\centering
\begin{subfigure}[t]{0.33\linewidth}
		\centering
		\includegraphics[width=\linewidth]{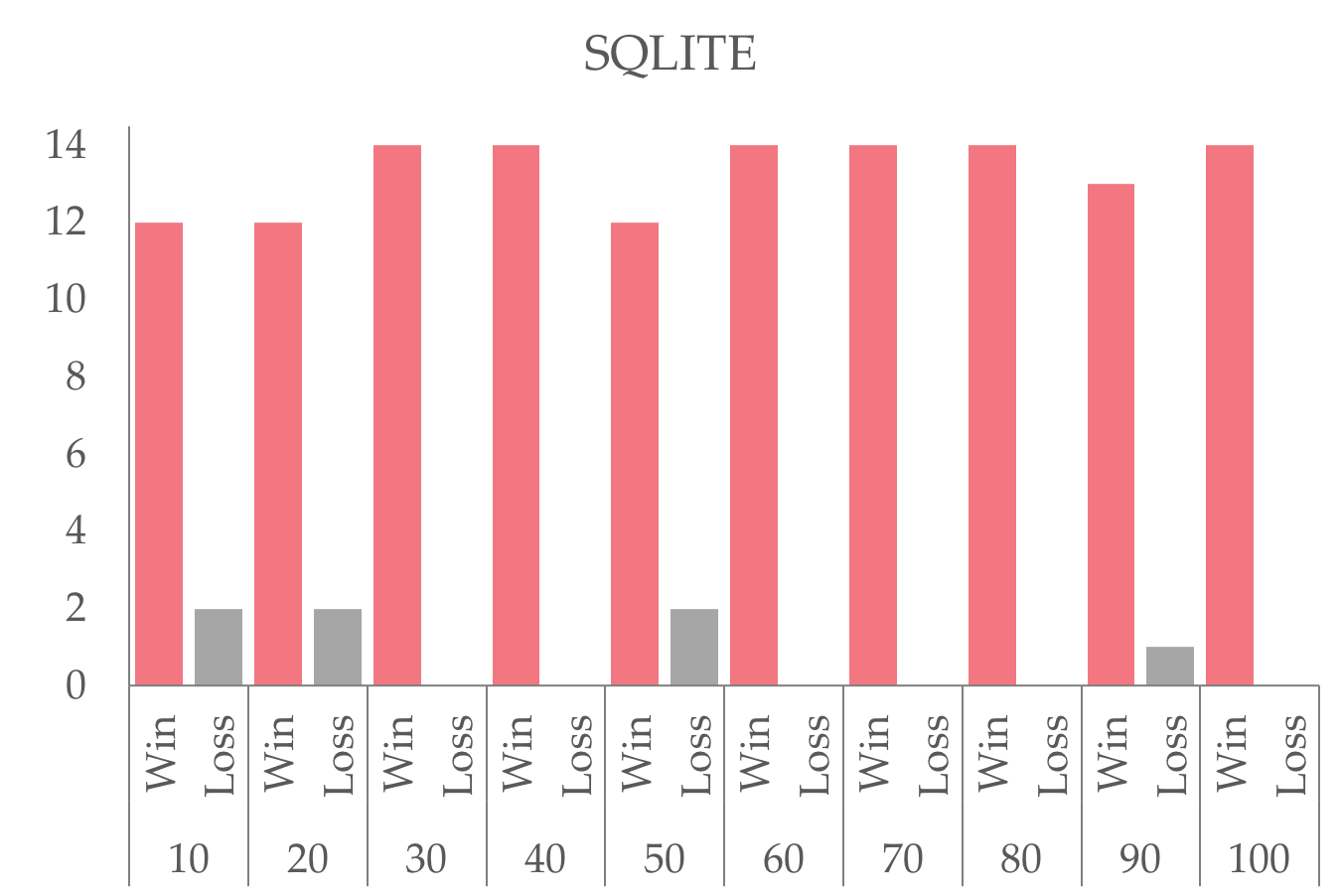}
\end{subfigure}%
\begin{subfigure}[t]{0.33\linewidth}
		\centering
		\includegraphics[width=\linewidth]{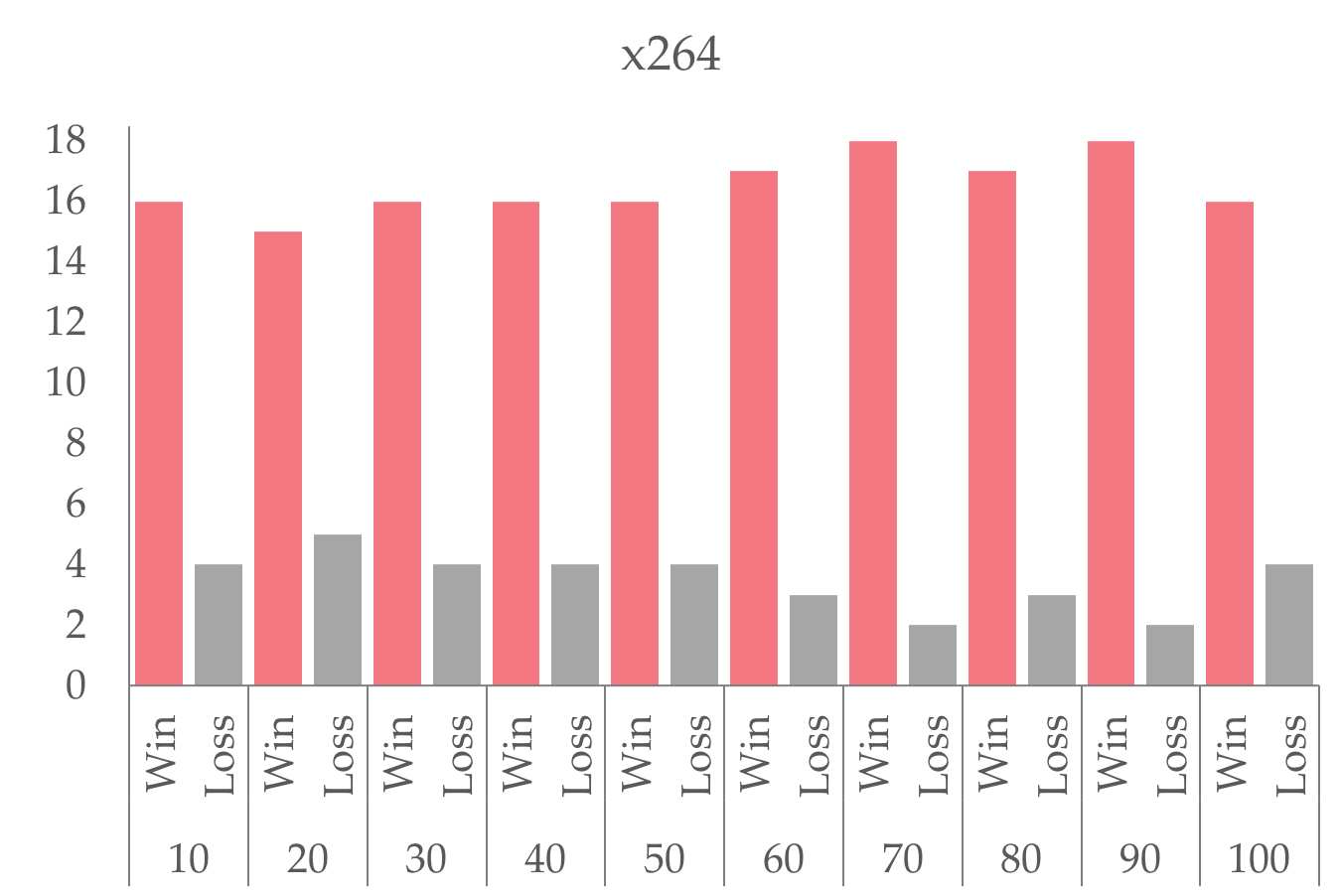}
\end{subfigure}%
\begin{subfigure}[t]{0.33\linewidth}
		\centering
		\includegraphics[width=\linewidth]{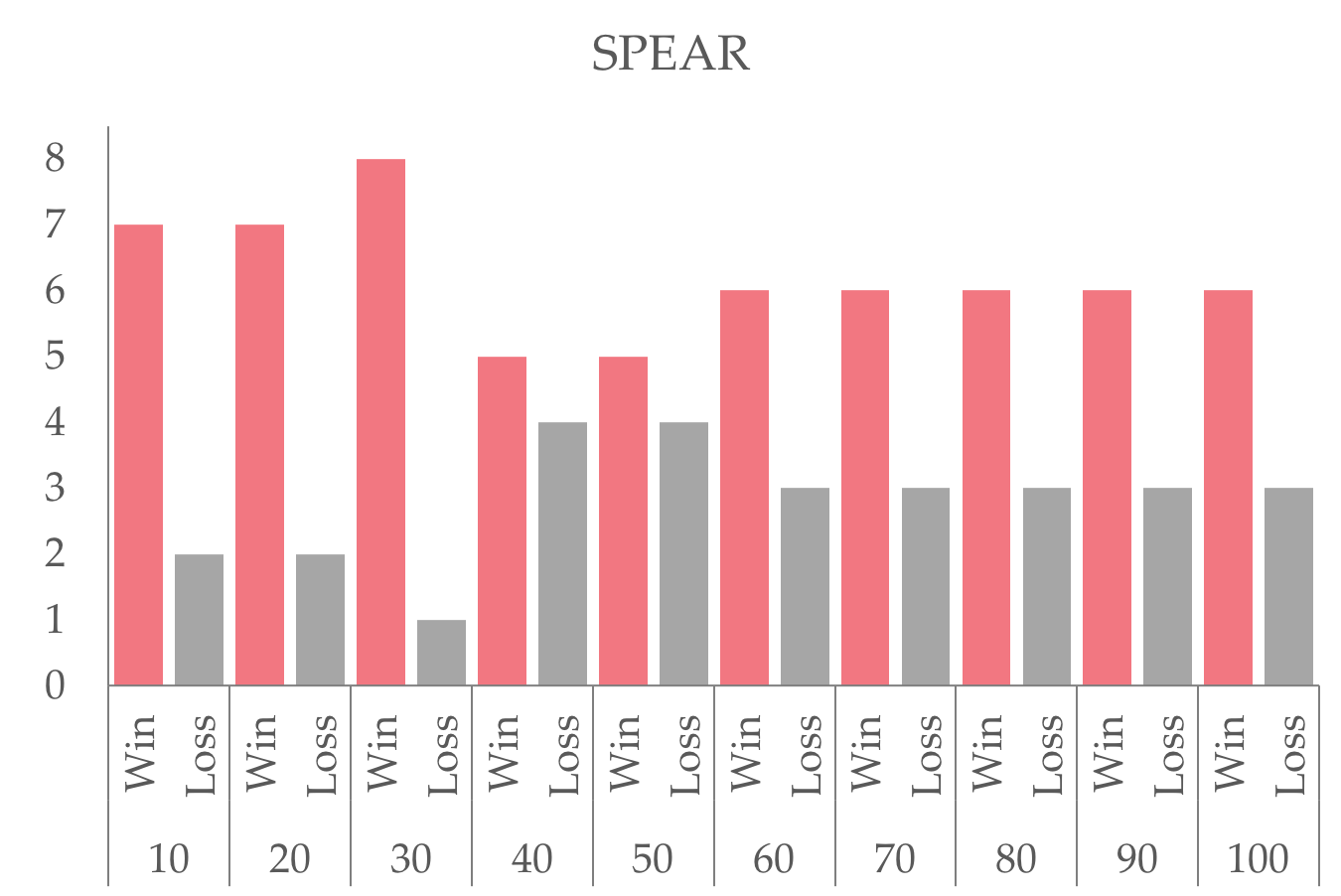}
\end{subfigure}\\[0.25cm]
\begin{subfigure}[t]{0.33\linewidth}
		\centering
		\includegraphics[width=\linewidth]{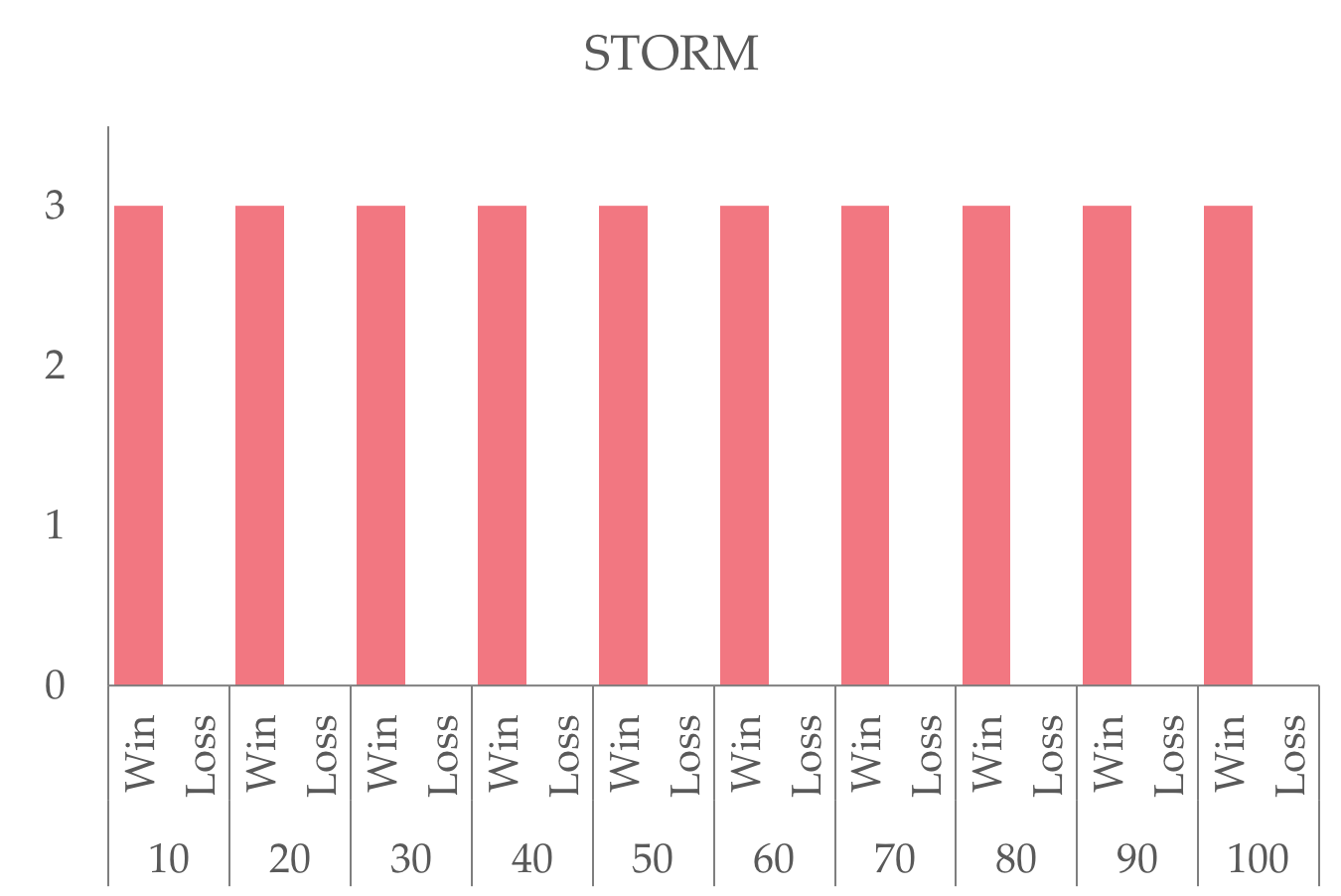}
\end{subfigure}\hspace{12pt}
\begin{subfigure}[t]{0.33\linewidth}
		\centering
		\includegraphics[width=\linewidth]{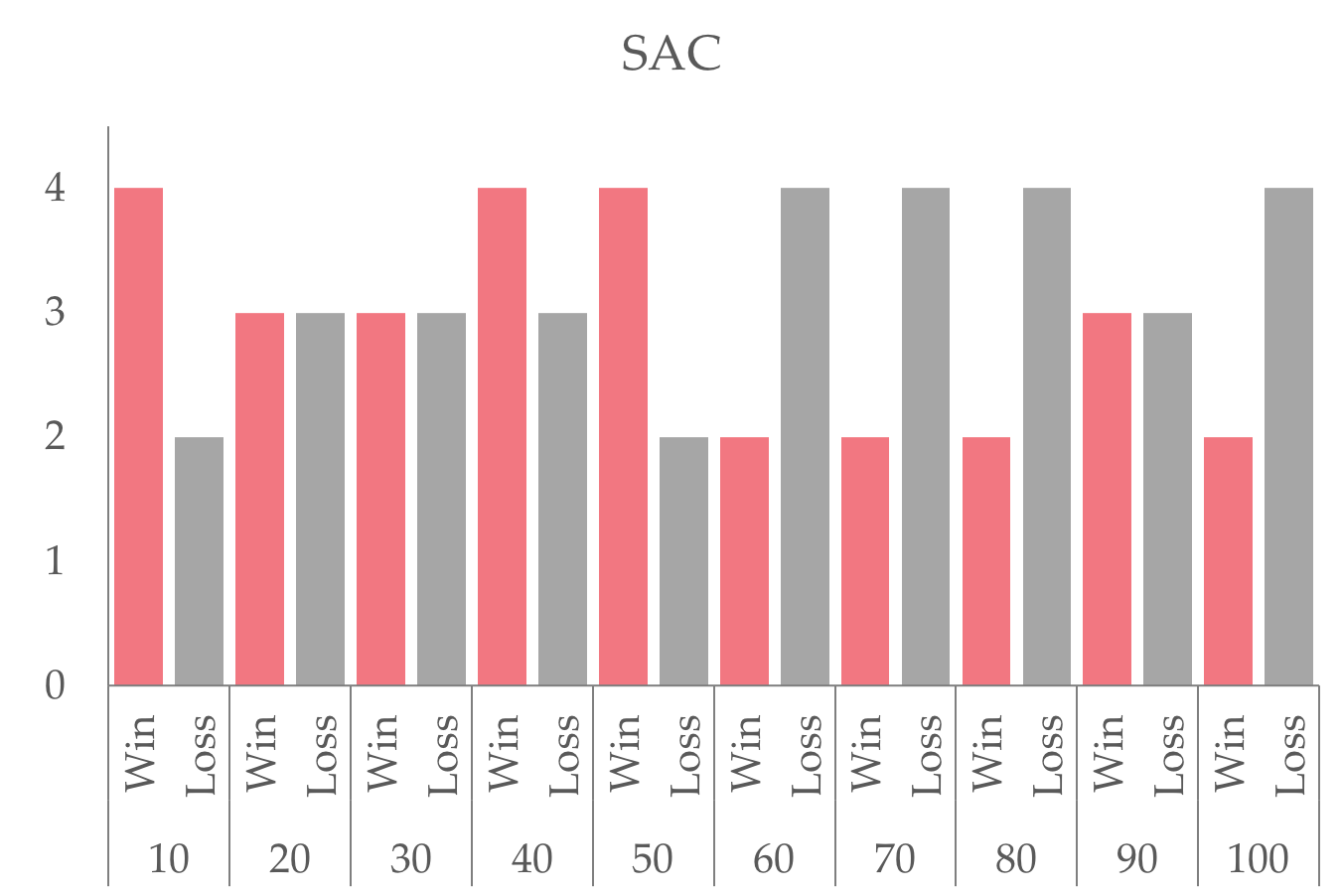}
\end{subfigure}\\[0.25cm]
	\caption{{\small Win/Loss analysis of learning from the bellwether environment and target environment using Scott Knott. The x-axis represents the percentage of data used to build a model and y-axis is the count.
	BEETLE wins in all models except for SAC-- and there only there when we have measured 50+\% of the data.}}
	\label{fig:rq2_1}
\end{figure*}

%% file: RQ4.tex
\begin{figure}[h!]
{\small
    \begin{minipage}[]{\linewidth}
    \textbf{{\sc Sac}}\\[0.01cm]
    \resizebox{\linewidth}{!}{%
    \begin{tabular}{|llrrc|}
    \arrayrulecolor{lightgray}
    \rowcolor{lightgray}{\small \textbf{Rank}} & {\small\textbf{Learner}} & {\small \textbf{Median}} & {\small\textbf{IQR}} & \\\hline  
    \rowcolor{lightergray}  1 &       Jamshidi et al.~\cite{jamshidi2017transfer} &    1.58  &  5.39 & \quart{0}{4}{0}{0} \\\hline
      2 &     Baseline &    6.89  &  99.1 & \quart{0}{79}{5}{0} \\
      2 &     Valov et al.~\cite{valov2017transferring} &    6.99  &  99.24 & \quart{0}{79}{6}{0} \\
    \hline \end{tabular}}\\[0.01cm]
    \textbf{{\sc Spear}}\\[0.01cm]
    \resizebox{\linewidth}{!}{%
    \begin{tabular}{|llrrc|}
    \arrayrulecolor{lightgray}
    \rowcolor{lightgray}{\small \textbf{Rank}} & {\small\textbf{Learner}} & {\small \textbf{Median}} & {\small\textbf{IQR}} & \\\hline  
    \rowcolor{lightergray}  1 &       Jamshidi et al.~\cite{jamshidi2017transfer} &    0.70  &  1.29 & \quart{0}{0}{3}{0}\\
    \rowcolor{lightergray}1 &     BEETLE &    0.79  &  1.40 & \quart{0}{0}{3}{0} \\
    \rowcolor{lightergray}1 &     Valov et al.~\cite{valov2017transferring} &    1.11  &  1.98 & \quart{0}{2}{4}{0} \\
    \hline\end{tabular}}\\[0.01cm]
    \textbf{{\sc SQLite}}\\[0.05cm]
    \resizebox{\linewidth}{!}{
    \begin{tabular}{|llrrc|}
    \arrayrulecolor{lightgray}
    \rowcolor{lightgray}{\small \textbf{Rank}} & {\small\textbf{Learner}} & {\small \textbf{Median}} & {\small\textbf{IQR}} & \\\hline  
    \rowcolor{lightergray}  1 &     BEETLE &    5.41  &  9.28 & \quart{2}{10}{5}{0} \\\hline
      2 &     Valov et al.~\cite{valov2017transferring} &    6.96  &  12.91 & \quart{3}{15}{6}{0} \\
      3 &     Jamshidi et al.~\cite{jamshidi2017transfer}   &    18.51  &  50.85 & \quart{2}{50}{18}{0} \\
    \hline \end{tabular}}\\[0.01cm]
    \textbf{{\sc Storm}}\\[0.01cm]
    \resizebox{\linewidth}{!}{%
    \begin{tabular}{|llrrc|}
    
    \arrayrulecolor{lightgray}
    \rowcolor{lightgray}{\small \textbf{Rank}} & {\small\textbf{Learner}} & {\small \textbf{Median}} & {\small\textbf{IQR}} & \\\hline  
    \rowcolor{lightergray}    1 &     BEETLE &    0.04  &  0.06 & \quart{0}{0}{0}{0} \\\hline
      1 &       Jamshidi et al.~\cite{jamshidi2017transfer} &    0.86  &  20.69 & \quart{0}{21}{1}{0} \\
      2 &     Valov et al.~\cite{valov2017transferring} &    2.47  &  53.98 & \quart{0}{54}{4}{0} \\
    \hline \end{tabular}}\\[0.01cm]
    \textbf{{\sc x264}}\\[0.01cm]
    \resizebox{\linewidth}{!}{%
    \begin{tabular}{|llrrc|}
    \arrayrulecolor{lightgray}
    \rowcolor{lightgray}{\small \textbf{Rank}} & {\small\textbf{Learner}} & {\small \textbf{Median}} & {\small\textbf{IQR}} & \\\hline  
    \rowcolor{lightergray}  1 &     BEETLE &    8.67  &  27.01 & \quart{2}{31}{8}{0} \\\hline
      2 &     Valov et al.~\cite{valov2017transferring} &    16.99  &  41.24 & \quart{5}{47}{17}{0} \\
      3 &       Jamshidi et al.~\cite{jamshidi2017transfer} &    43.58  &  28.39 & \quart{34}{48}{44}{0} \\
    \hline \end{tabular}}
    \end{minipage}}
    \caption{{\small Comparison between state-of-the-art transfer learners and BEETLE. The best transfer learner is shaded \colorbox{lightergray}{gray}.}}
    \label{fig:rq4}
    \end{figure}
    
    \begin{figure}[t]
        \centering
        \includegraphics[width=0.95\linewidth]{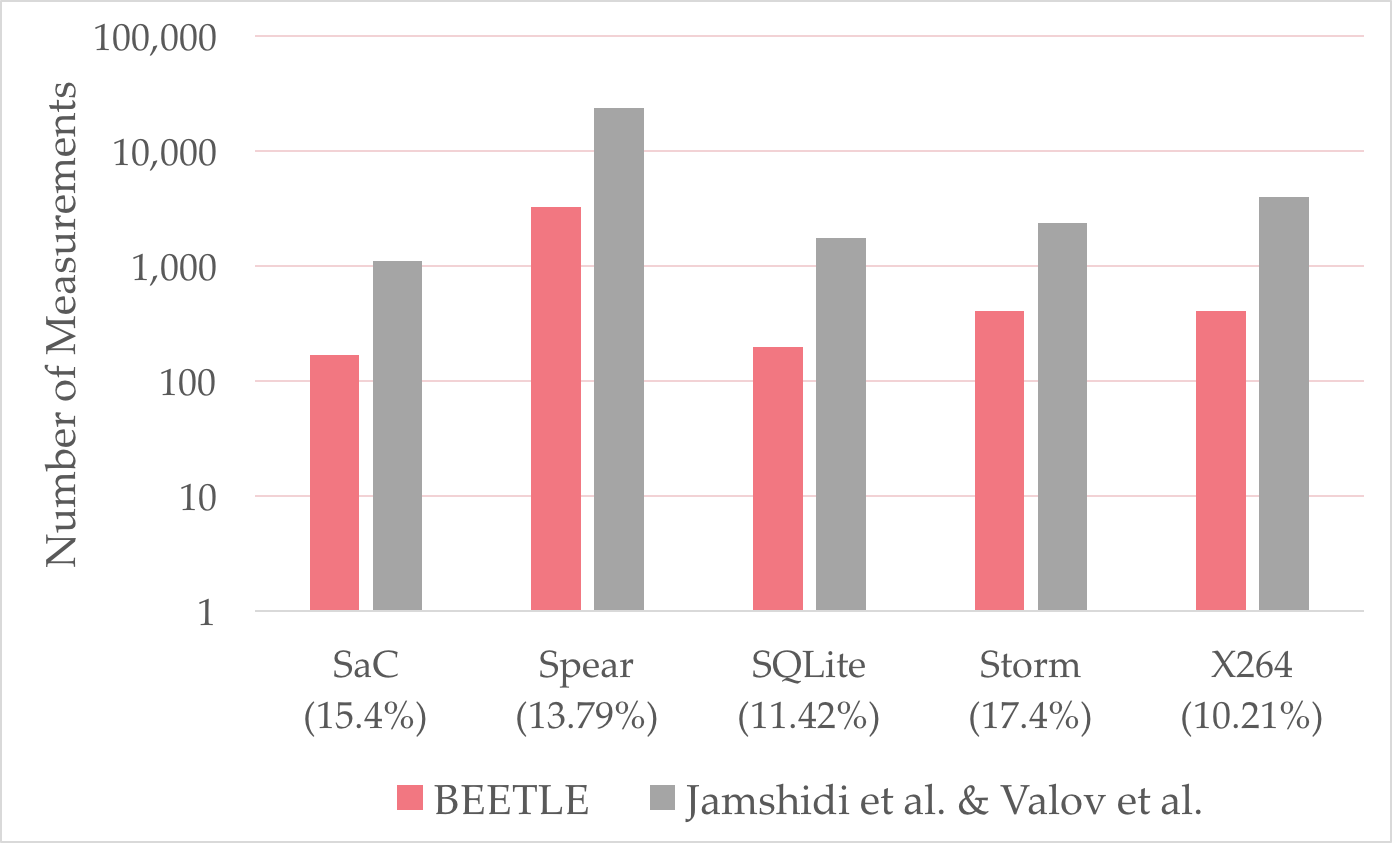}
        \caption{{\small BEETLE v/s  state-of-the-art transfer learners. The numbers in parenthesis represent the numbers of measurements BEETLE uses in comparison to the  state-of-the-art learners.}}
        \label{fig:rq4_measurement}

\end{figure}